\documentclass[prx,twocolumn,showpacs,preprintnumbers,amsmath,amssymb]{revtex4-1}

\usepackage{graphicx}
\usepackage{dcolumn}
\usepackage{bm}

\begin{document}

\title{Diffraction catastrophes threaded by quantized vortex skeletons caused by atom-optical aberrations induced in trapped Bose-Einstein condensates}
\author{T. P. Simula$^1$, T. C. Petersen$^{1,2}$ and D. M. Paganin$^1$}
\affiliation{$^1$School of Physics, Monash University, Victoria 3800, Australia}
\affiliation{$^2$Monash Centre for Electron Microscopy, Monash University, Victoria 3800, Australia }

\begin{abstract}
We propose a nonlinear atom-optics experiment to create diffraction catastrophes threaded by quantized vortex skeletons in Bose--Einstein condensed matter waves. We show how atom-optical aberrations induced in trapped Bose--Einstein condensates evolve into specific caustic structures due to imperfect focusing. Vortex skeletons, whose cross-sections are staggered vortex lattices, are observed to nucleate inside the universal diffraction catastrophes. Our observations shed further light on the structure and dynamics of Bose-novae and suggest applications of matter wave diffraction catastrophes, including detection of the order parameter pairing symmetry in cold gas experiments.
\end{abstract}

\maketitle

\section{Introduction}
Focusing is a generic phenomenon occuring in nature at all scales. Gravitational collapse leading to supernovae and black holes \cite{Baade1934a}, sonoluminesence from imploding air bubbles \cite{Brenner2002a}, image formation on the retina of a human eye and Bose-novae of collapsing Bose--Einstein condensates \cite{Sackett199a,Donley2001a,Lahaye2008a,Aikawa2012a} all involve focusing of light or matter into a small volume of space. Focusing is pivotal to many technological applications such as optical and electron microscopy, gravitational lensing and photography.

Natural focusing is rarely perfect, lens imperfections causing the well known optical aberrations such as astigmatism and coma. Imperfect lensing explodes a point focus, yielding extended caustics occupying a focal volume. The internal structure of such caustics can be described by the diffraction theory of aberrations \cite{Wolf1951a} and catastrophe theory \cite{Thom1975a,Arnold1984a,Berry1980a,Berry1979a}. Remarkably, on closer inspection such caustics, which in ray optics correspond to infinite intensities, turn out to be perforated by a skeleton of quantized vortices and antivortices whose void cores define lines of strictly zero intensity \cite{Berry1998a}. In the spirit of particle--wave duality, Berry has put forth a conjecture of a caustic--vortex duality whereby the infinite intensities in the caustics are complementary to the zero intensity vortex cores \cite{Berry1981a}. 

The rich structure of quantized vortices which emerge inside caustics can be understood in terms of multi-wave interference phenomena where three \cite{Pearcey1946a} or more waves interfere destructively to produce lattices of quantized phase singularities \cite{Nicholls1987a,Oholleran2006a,Paganin2006a,Scherer2007a,Ruben2008a,Ruben2010a,Simula2011a}. Such nodal lines generically appear in systems described by complex valued wavefunctions. Optical vortices \cite{Nye1974a} and recently observed vortex knots \cite{Dennis2010a} together with their matter wave cousins, electron vortices \cite{Uchida2010a,Verbeeck2010a,McMorran2011a,Petersen2013a} and electron vortex loops \cite{Petersen2013a}, are examples of systems where coherent linear superposition of multiple wave trains yield nodal lines in complex wavefunctions. In addition to these linear systems, quantized vorticity is intimately connected with the phenomenon of superfluidity observed for example in He II \cite{Donnelly1991a} and Bose--Einstein condensed atomic gases \cite{Fetter2009a}. In superfluids, nonlinearities due to particle interactions are an inherent property of the system and the concept of a quantized vortex acquires perhaps its closest analog to the intuitive picture of a classical whirlpool.

\begin{figure}
\includegraphics[width=0.9\columnwidth]{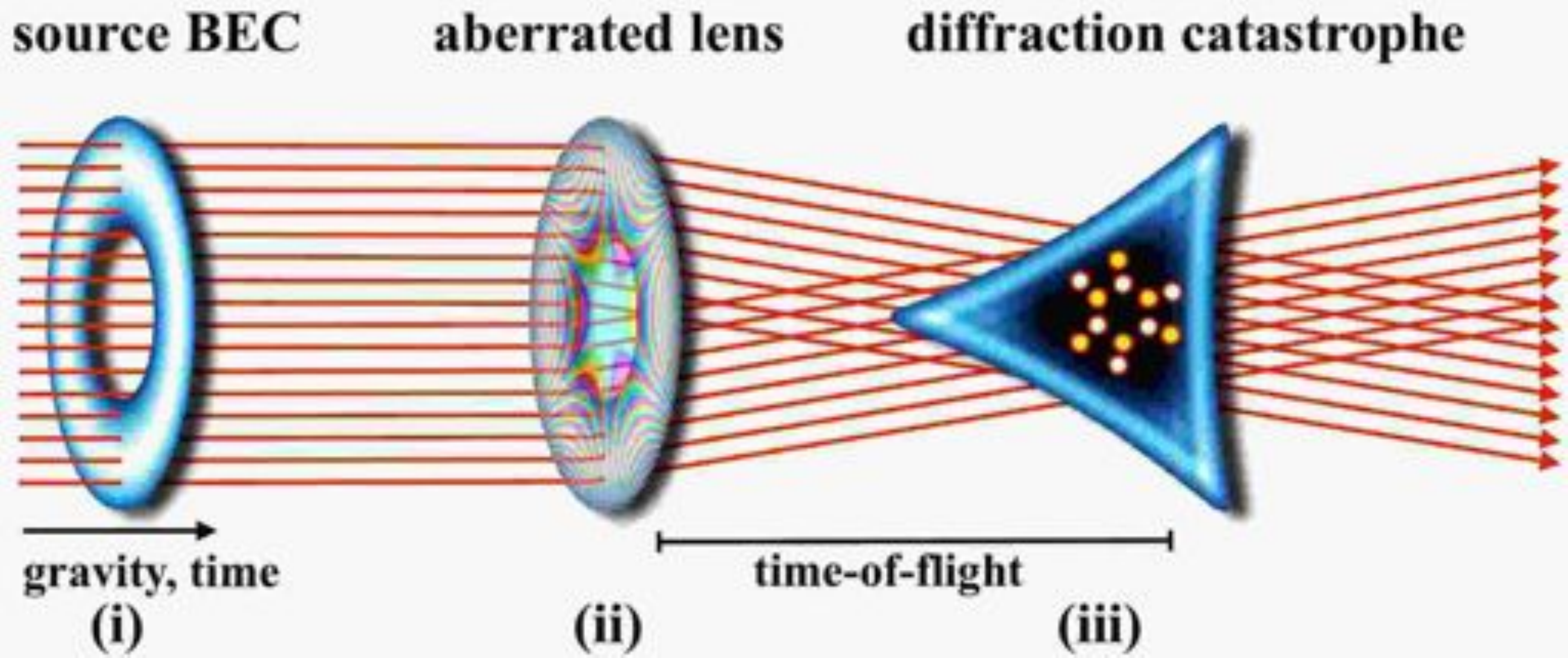}
\caption{(Color) Schematic of proposed singular atom-optics experiment. (i) Bose--Einstein condensate in ground state of a ring trap is aberrated by an atom-optical lens (ii) applying a perturbation to the trapping potential which imprints a phase profile to the condensate wavefunction affecting its momentum distribution. Such lensing perturbations could be achieved using spatial light modulators or time averaged scanning light-shift potentials. The diffraction catastrophe and the quantized vortex skeleton emerge during the time-of-flight due to multi-wave interference. The matter wave probability density can be directly captured on a charge-coupled device at a desired propagation distance. The colored circles in (iii) illustrate the cross-section of the vortex skeleton nucleated inside the diffraction catastrophe with a staggered vortex lattice of opposite signs of quantized circulation. } 
\label{fig1}
\end{figure}

Aberrated lensing, caustics, diffraction catastrophes and their connection to quantized vortices have been observed in light optics and more recently in electron matter waves \cite{Petersen2013a}. Similar experiments focusing on matter wave lensing in superfluids are within reach of existing technology. However, a notable difference between photons or electrons that are used for microscopy and the condensates of cold atoms is that the latter are tuneably self-interacting and result in nonlinear atom-optics effects such as four-wave mixing \cite{Deng1999a}. 

Here we propose diffractive singular atom-optics experiments deploying particle interactions of the Bose--Einstein condensates to create and observe the internal structure of caustics and diffraction catastrophes which are punctured by quantized vortex skeletons. We show how nonlinear atom-atom interactions can induce drastic changes to the structure of such quantized vortex skeletons due to self-focusing. We consider a coherent ensemble of interacting Bose--Einstein condensed atoms, lensed both by the external potentials used for trapping and manipulating the atoms and the particle interactions. Such lens imperfections result in formation of condensate matter wave caustics \cite{Chalker2009a,Efremidis2013a}. The generic scenario considered in this paper, where matter wave diffraction catastrophes spawn formation of quantized vortex skeletons in aberrated Bose--Einstein condensates, is illustrated by the schematic Fig.~\ref{fig1}.


\section{Model}
We model the structure and dynamics of weakly interacting Bose--Einstein condensates using the Gross--Pitaevskii Hamiltonian 
$H = -\frac{\hbar^2\nabla^2}{2m}+V_{\rm ext}({\bf r},t)+V_{\rm int}({\bf r},t) $, where $\hbar$ is Planck's constant, $m$ is particle mass and $V_{\rm ext}({\bf r},t)$ is an external time-dependent potential used to confine and manipulate the atoms, explicitly accounting for the aberrating imperfections such as the anisotropy of the trapping potential \cite{Pitaevskii1961a,Gross1961a}. The particle interactions yield the potential $V_{\rm int}({\bf r},t)$ which, unless otherwise stated, is a contact interaction $V_{\rm int}({\bf r},t)=g|\psi({\bf r},t)|^2 $ where $g=4\pi\hbar^2a/m$ is the interaction coupling constant proportional to the $s$-wave scattering length $a$, and the strength and sign of the atom-atom interactions $g$ can be routinely controlled using a Feshbach resonance technique \cite{Kohler2006a}.

\begin{figure}
\includegraphics[width=1\columnwidth]{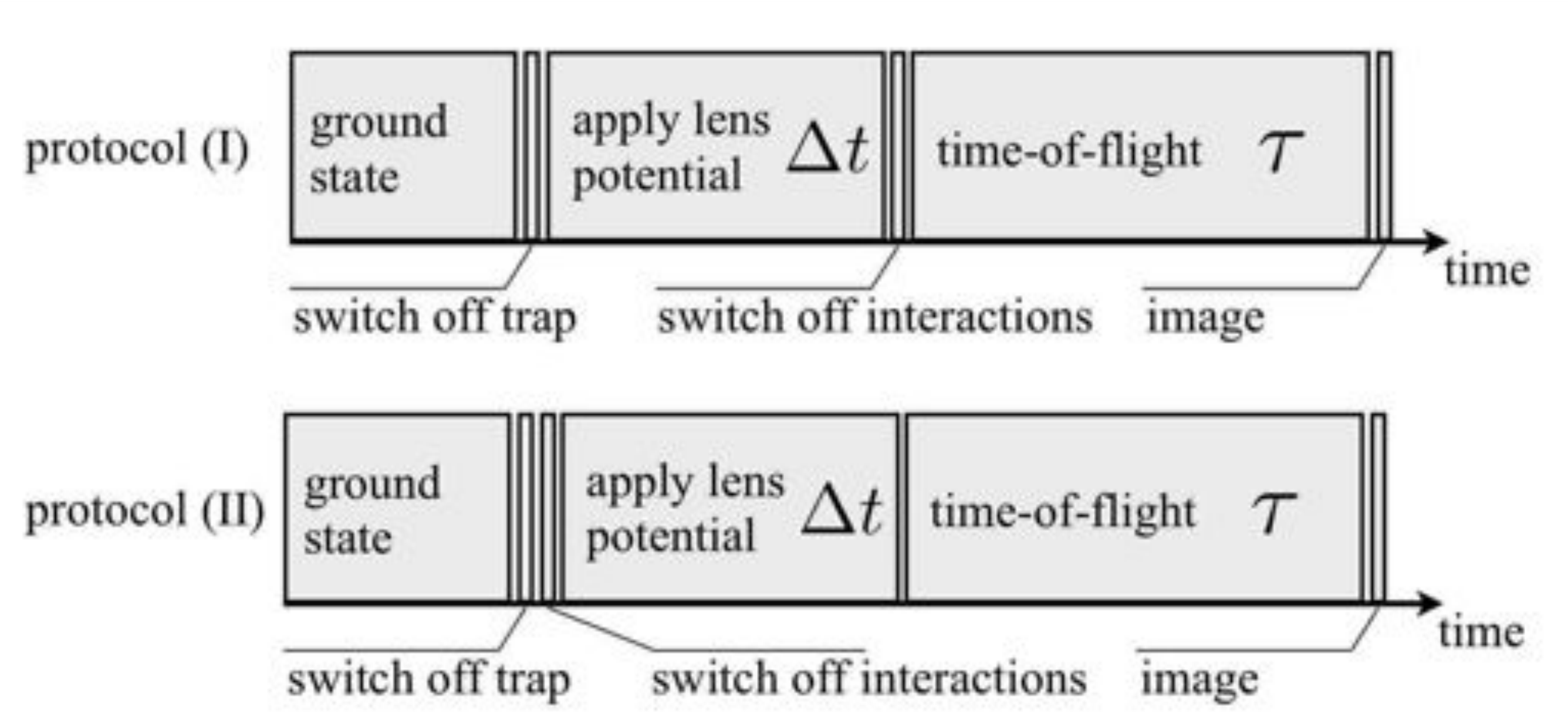}
\caption{Timing sequence of two matter wave lensing protocols (I) and (II).} 
\label{fig6}
\end{figure}

Our typical numerical experiment begins with a Bose--Einstein condensate confined in a toroidal trap $V_{\rm ext}({\bf r},t=0) = V_{\rm{osc}}({\bf r}) + V_{\rm{LG}}({\bf r})$, where the harmonic oscillator potential $V_{\rm{osc}}({\bf r})=m(\omega_x^2x^2+\omega_y^2y^2+\omega_z^2z^2)/2$ with $\omega_i$ the Cartesian frequencies and the Mexican hat structure is achieved using a red-detuned Laguerre--Gauss laser mode to result in a potential $V_{\rm{LG}}({\bf r})=  -500 (x^2+y^2) /\sigma^2 e^{-2(x^2+y^2)/\sigma^2}\hbar\omega_\perp$, where $\sigma = 20 a_{\rm osc}$, with $a_{\rm osc}=\sqrt{\hbar/m\omega_\perp}$ being the harmonic oscillator length \cite{Wright2000a,Ryu2007a,Ramanathan2011a,Simula2008a}. We consider cylindrically symmetric initial trap configurations $\omega_\perp\equiv\omega_x=\omega_y$ with $\omega_z=5\omega_\perp$ and interaction strength $g'=gN/\hbar\omega_\perp a_{\rm osc}^3=5000$, where $N$ is the number of particles in the condensate. In the ground state of this potential the condensate takes the shape of the usual Thomas--Fermi doughnut and the phase map $S({\bf r})=\arg(\psi({\bf r},t))$ is spatially constant throughout the condensate. The choice of these parameters correspond to typical experimental values \cite{Ramanathan2011a}.

The motivation for using ring trapped condensates is to remove the high intensity of atoms from within the interior of the caustic structures, which would otherwise hinder the formation of vortices inside the diffraction catastrophes. Moreover, the stability of the caustic structures increases as the width of the toroid is reduced and in the limit of infinitely narrow condensate toroids the caustics approach propagation invariance. The form of our toroidal condensate is further motivated by recent experiments which utilized such a trapping geometry to study persistent currents in Bose--Einstein condensates \cite{Ryu2007a,Ramanathan2011a}. In the context of this work, such trapping potentials are apt for creating and experimentally observing  vortex skeletons, which nucleate within the diffraction catastrophes of aberrated Bose--Einstein condensates. 

Consider an externally applied potential of the form
$V_{\rm lens}({\bf r})=\sum_{n=0}^\infty \sum_{m=-n}^n\beta(n,m) {\mathcal Z}_n^m$
where $\beta(n,m)$ are real numbers and ${\mathcal Z}_n^m$ is a Zernike polynomial indexed by $(n,m)$ \cite{Born1999a}. When applied for a short duration $\Delta t $ (in comparison to other relevant time scales in the problem) this potential corresponds to the atom-optic analog of a thin lens operating in the Raman--Nath regime of diffraction \cite{Meystre2001a,Simula2007a,Chu2008a}. More generally, when applied for longer durations, both the phase and intensity of the matter wave will evolve substantially during the lensing. In either case, diffraction catastrophes are expected to develop. Each Zernike polynomial ${\mathcal Z}_n^m$ corresponds to a different kind of lens aberration such as astigmatism or coma. The effect of such a potential pulse on a trapped ground state Bose--Einstein condensate is to imprint a complex phase to the macroscopic wavefunction describing the condensate. Since the momentum of the atoms is proportional to the gradient of this phase, it also amounts to imparting a momentum distribution to the condensate, specified by the chosen aberration potential, see Fig.~\ref{fig1} stage (ii).


\begin{figure}
\includegraphics[width=1\columnwidth]{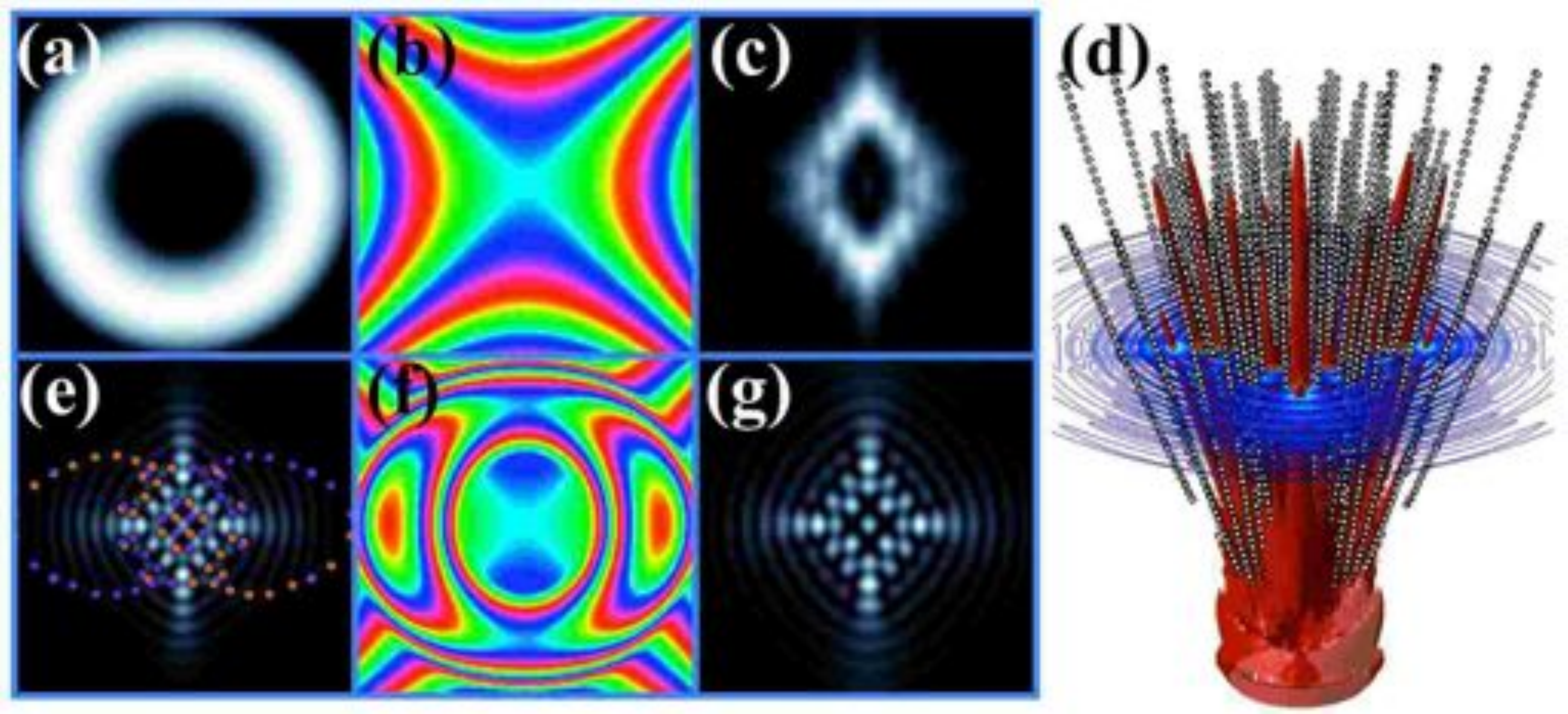}
\caption{(Color) Astigmatism induced ${\mathcal Z}_2^2$ $d$-wave diffraction catastrophes in Bose--Einstein condensate matter waves. Frame (a) shows a slice of the condensate density immediately after it has passed through the lens. Frames (b) and (f) show the corresponding phase maps obtained using lensing methods (II) and (I), respectively, and frames (c) and (g) show the respective condensate densities after ballistic evolution. Frame (e) shows the staggered vortex lattice (vortices and antivortices are marked using different colors) obtained by cutting through the vortex skeleton shown in (d). The fields of view of the images are (a) $12\times12$, (b) $12\times12$, (c) $100\times100$, (e) $50\times50$, (f) $12\times12$, and (g) $25\times25$, in units of $a_{\rm osc}$. The images are acquired after a ballistic expansion of a duration
(c) $\tau\omega_\perp=37.7$, (g) $\tau\omega_\perp=6.9$, and (e) $\tau\omega_\perp=11.9$. In frame (d), which shows the vortex skeleton and the surrounding isodensity surface, the time axis runs from zero (bottom) to $\tau\omega_\perp=18.9$ (top). The frame (e) corresponds to the contour shown in (d).} 
\label{fig2}
\end{figure}

Here we begin the matter wave lensing by switching off the external trapping potentials confining the ground state condensate. We thereafter consider two different experimental sequences illustrated in Fig.~\ref{fig6}: (I) apply the perturbing lens potential $V_{\rm lens}({\bf r})$ to the interacting condensate for duration $\Delta t$, after which a Feshbach resonance is used to tune the interaction strength to $g'=0$. The condensate is then left to evolve ballistically in the time-of-flight for a time $\tau$ before measuring the probability density distribution. In the second protocol (II) we tune close to $g'=0$ first at the same time when the trapping potentials are turned off and after this apply the lens aberration for duration $\Delta t$ before imaging after an additonal duration $\tau$ of time-of-flight. Note that setting the lens perturbation equal to the original trapping potential, which is tantamount to not switching the trap off in the first place, in the case (II) corresponds to a typical Bose-nova experiment where the condensate is focused in-trap due to the removal of the mean-field repulsion (or even changing its sign). A third option (III) would be to let the atoms interact throughout the experiment to observe how the nonlinear interactions cause bending of the atom trajectories for the whole duration of the time-of-flight, although the effect of these interactions would be rapidly washed away as the cloud is diluted during its free expansion. While experimentally straightforward, we do not pursue this third option further here.

For a non-interacting condensate the final state $\psi({\bf  r},t_f)$ of the condensate matter wave, after the ballistic time-of-flight evolution for the duration $\tau$, initially in a state $\psi({\bf  r},t_i)$, is obtained by integrating the Schr\"odinger equation
$
\psi({\bf r},t_f) = \mathcal{F}^{-1}\left\{e^{-i\hbar{\bf k}^2\tau/2m} \mathcal{F} \left (\psi({\bf  r},t_i)\right)\right\}
$
where $\mathcal{F}$ denotes a Fourier transform and ${\bf k}$ is a wavevector. 
For linear systems the initial state $\psi({\bf  r},t_i)\approx e^{-iV_{\rm lens}({\bf r})\Delta t/\hbar}\psi({\bf  r},0)$ can be calculated by propagating the ground state $\psi({\bf  r},0)$ by using the potential operator only. However, for interacting condensates the solution must be obtained by propagating the field using the full Hamiltonian including the self-interactions. We obtain the initial states  $\psi({\bf  r},t_i)$ by solving the three-dimensional time-dependent Gross--Pitaevskii equation until the particle interactions and all trapping potentials are turned off. The Fourier transformed condensate wavefunction is then multiplied by a free propagator and back transformed to real space to obtain the spatial condensate density corresponding to the experimentally measurable probability density of atoms.

\begin{figure}
\includegraphics[width=0.8\columnwidth]{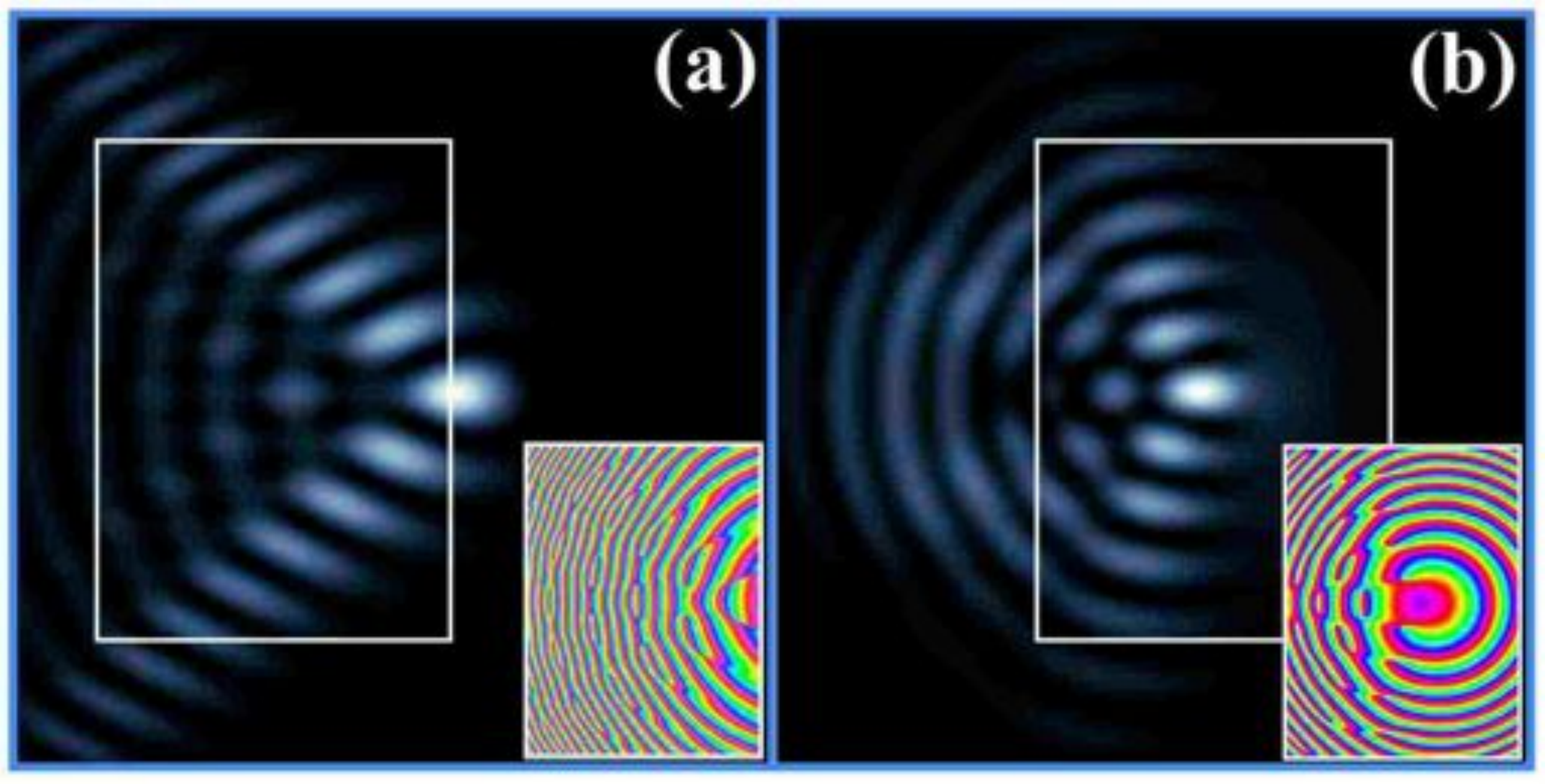}
\caption{(Color) Coma induced ${\mathcal Z}_3^1$ hyperbolic umbilic diffraction catastrophes in ring trapped Bose--Einstein condensate matter waves. The subplots in the lower right corners are phase maps corresponding to the boxed regions. The horizontal and vertical axes of the main images run from $-80 a_{\rm osc}$ to $40 a_{\rm osc}$ and from $-60 a_{\rm osc}$ to $60 a_{\rm osc}$, respectively.} 
\label{coma}
\end{figure}

\section{Results}

\subsection{Astigmatism aberration}
Figure \ref{fig2} (a) shows the condensate density $|\psi(z=0)|^2$ immediately after application of an astigmatic $d$-wave lens perturbation $V_{\rm lens}({\bf r})= 0.08 (x^2 -  y^2) \hbar \omega_\perp/a_{\rm osc}^2\propto {\mathcal Z}_2^2$ for the duration $\Delta t\omega_\perp =0.005$.
The unbroken cylindrical symmetry in Fig.~\ref{fig2} (a) shows that the aberrating lens perturbation was applied for a sufficiently short duration to leave the condensate probability density undisturbed. However, the phase of the wavefunction, shown in (b) and (f), and which determines the momentum distribution of the atoms, is drastically different depending on whether method (II) or (I)  is used, respectively. This is because the nonlinear particle interactions cause intrinsic lensing or self-focusing, in addition to the externally applied lens potential. This nonlinear effect becomes readily observable in the far field images of the condensate density. Frames (c) and (g) display the particle probability density after $\tau\omega_\perp=37.7$ and $\tau\omega_\perp=6.9$ of time-of-flight showing how the respective momentum distributions in (b) and (f) are sources of different internal diffraction detail of the caustics. Frame (d) shows the full quantized vortex skeleton (white dotted lines) which supports the caustic structure, together with an isodensity surface illustrating the caustic flesh. The caustic body has been cut through with a contour plot and the corresponding caustic cross-section, which reveals the staggered lattice of quantized vortices and antivortices, is shown in (e). Movies showing the evolution of these caustics are in the Supplemental Material \cite{Supplement}.

The ``salmiakki" shaped caustics in Fig.~\ref{fig2} embody the doubly folded manifolds with cusps in the corners of the caustic. Such flared rhombi are the telltale signature of astigmated lensing. The structure in Fig.~\ref{fig2}~(g) reveals that the combination of astigmatism and intrinsic nonlinear lensing produces a caustic similar to an astigmated axicon \cite{Tanaka2000a}. The toroidal trapping of the condensates enhances vortex production inside the caustics when compared with more conventional harmonically trapped systems.

\begin{figure}
\includegraphics[width=0.9\columnwidth]{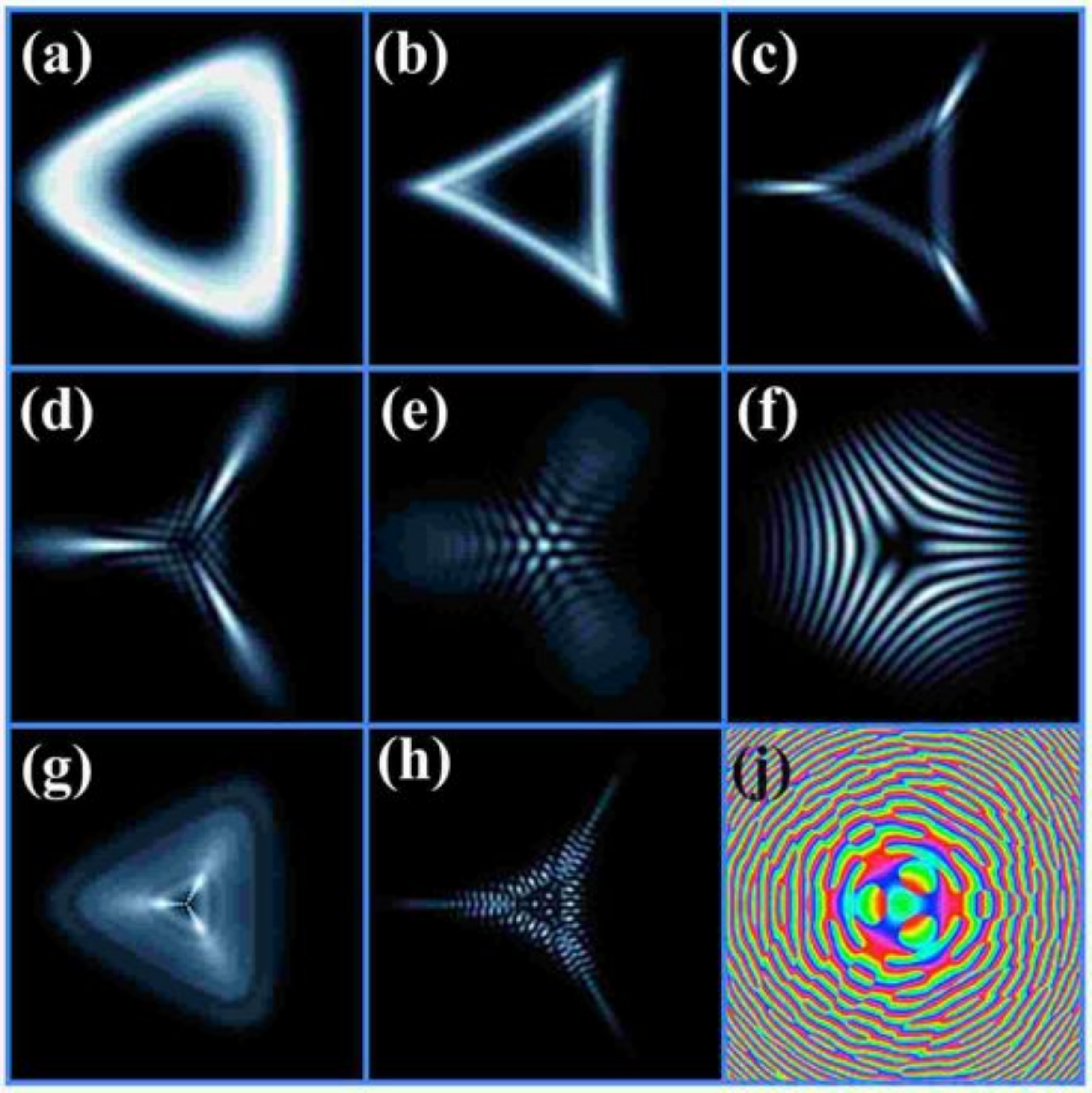}
\caption{(Color) Elliptic umbilic ${\mathcal Z}_3^3$ $f$-wave diffraction catastrophes in Bose--Einstein condensate matter waves. Frames (a)-(f) show snapshots of a diffraction catastrophe induced by a trefoil aberration using method (II). Frames (g) and (h) are obtained using method (I)  and frame (j) is a phase map corresponding to the central region of frame (h). Note that the phase map is multivalued at points defining the cores of the quantized vortices. The field of view of the images (a)-(j) are $15\times15$, $20\times20$, $20\times20$, $20\times20$, $25\times25$, $100\times100$, $25\times25$, $180\times180$, and $50\times50$, in units of $a_{\rm osc}$, respectively. The respective frames correspond to $\tau\omega_\perp = 0.75, 1.5, 2.5, 3.5, 6.0, 25, 1.0, 25$, and $25$ of ballistic expansion.} 
\label{fig4}
\end{figure}

\begin{figure}[!t]
\includegraphics[width=0.9\columnwidth]{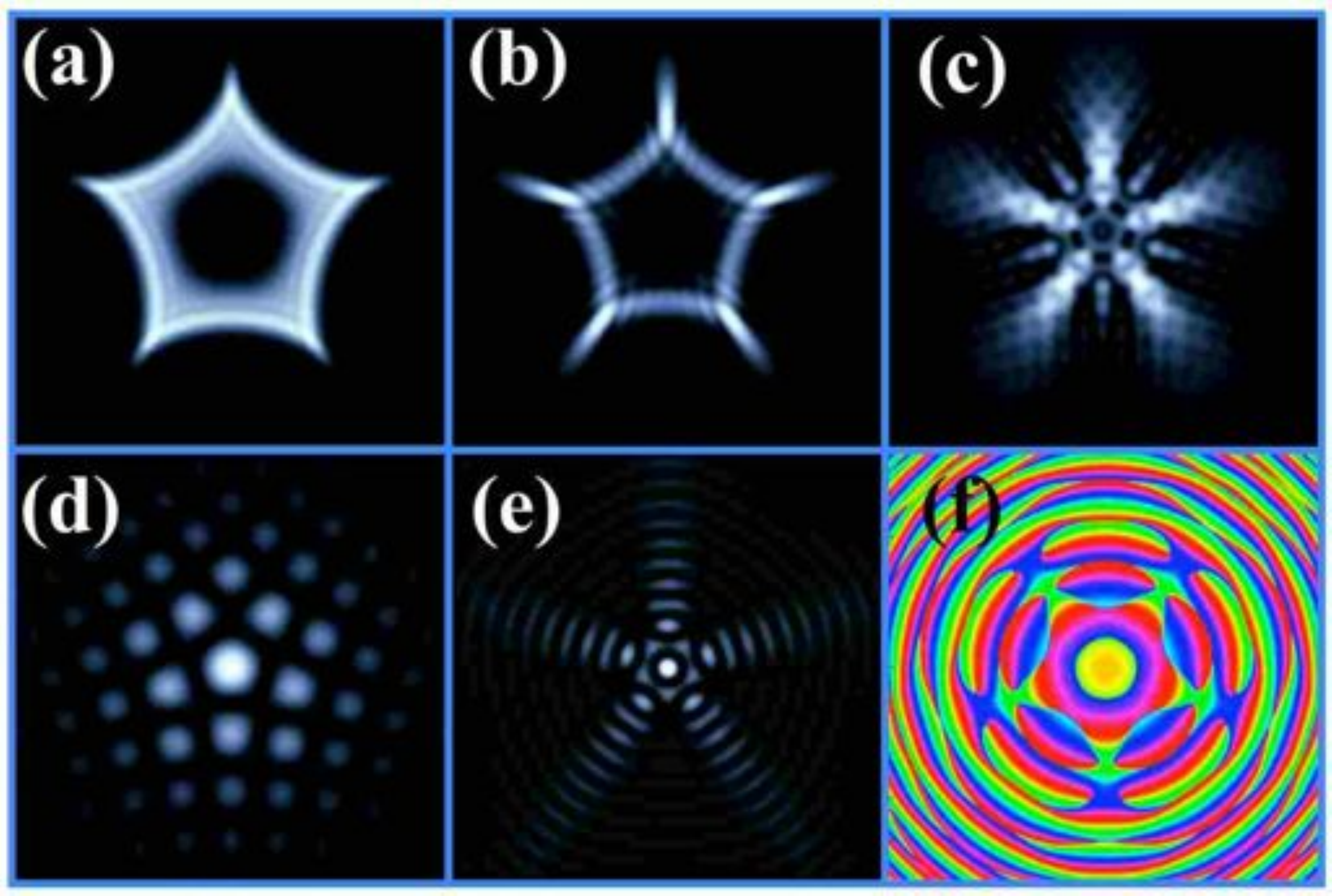}
\caption{(Color) Pentafoil ${\mathcal Z}_5^5$ $h$-wave caustics in Bose--Einstein condensate matter waves. Frames (a)-(d) show snapshots of a diffraction catastrophe induced by a pentafoil aberration using method (II) and the frames (e) and (f) are obtained using method (I). The field of view of the images (a)-(f) are $20\times20$, $20\times20$, $20\times20$, $20\times20$, $100\times100$, $50\times50$, and $25\times25$, in units of $a_{\rm osc}$, respectively. The respective frames correspond to $\tau\omega_\perp = 1.0, 2.0, 5.0, 25, 12.6$, and $12.6$ of ballistic expansion. Frame (f) is a phase map corresponding to the central region of (e).} 
\label{fig5}
\end{figure}

\subsection{Coma aberration}
Figure \ref{coma} (a) and (b) illustrate the hyperbolic umbilic diffraction catastrophe of ring-trapped Bose--Einstein condensates for the coma aberration $V_{\rm lens}({\bf r})= (3x-12x (x^2+y^2)+10x(x^2+y^2)^2) / 10000\propto{\mathcal Z}_3^1$. The insets show the phase maps corresponding to the boxed regions in the main images. Method (II) produces a caustic structure well known from linear optics, whereas method (I)  displays slightly different vortex structure due to the nonlinear effects. For these cases the momentum imparted by the lens on the atoms is moderate but by using stronger lens perturbations more vortices and finer diffraction detail can be produced. The internal lensing in the coma case causes less dramatic changes to the diffraction catastrophe than the astigmatism because the symmetry of the coma aberration is closer to the cylindrical symmetry of the mean-field aberration. The evolution of the coma aberrated condensate is shown in the Supplemental Material \cite{Supplement}.

\subsection{Trefoil aberration}
Figures \ref{fig4}~(a)-(f) show a through focus series of images of a condensate particle density resulting from a trefoil f-wave lens perturbation of the form $V_{\rm lens}({\bf r})= 0.01(x^3 - 3xy^2)\hbar\omega_\perp/a_{\rm osc}^3\propto{\mathcal Z}_3^3$, applied for duration $\Delta t\omega_\perp=0.005$, using method (II) over consecutive ballistic expansion times, which portray sections of the elliptic umbilic diffraction catastrophe. Frames (g) and (h) show the condensate density after $\tau\omega_\perp=1.0$ and $\tau\omega_\perp=25$ of ballistic expansion, respectively, when method (I)  is used. Again, the nonlinear particle interactions induce enriched quantized vortex skeletons inside the diffraction catastrophes. Frame (j) shows the phase map corresponding to central region of the frame (h). Movies showing the evolution of the caustics are in the Supplemental Material \cite{Supplement}. 

\begin{figure*}
\includegraphics[width=1.8\columnwidth]{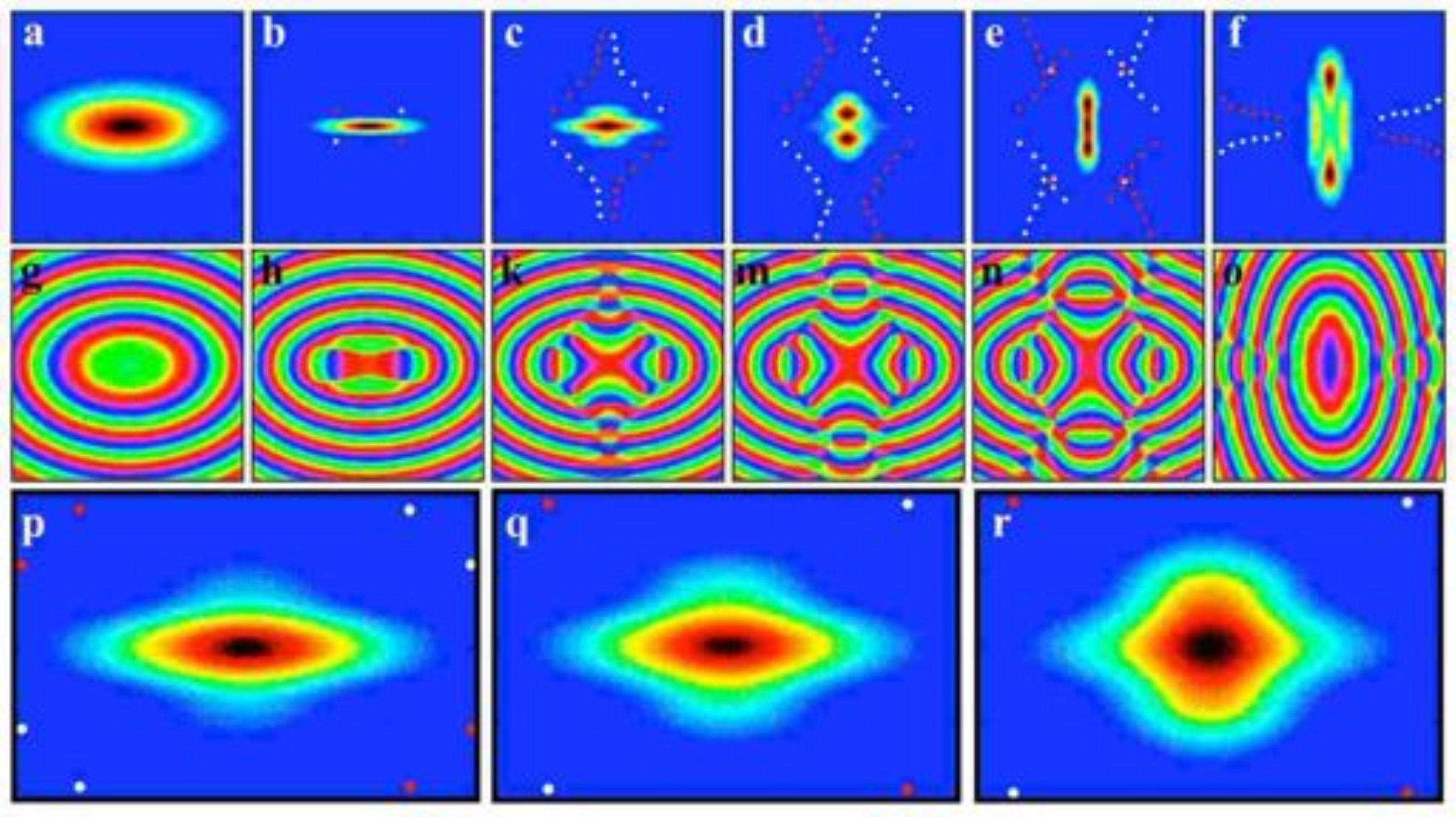}
\caption{(Color) Astigmatism induced caustic structures in anisotropic harmonically trapped Bose--Einstein condensate. Frames (a-f) are snapshots of the condensate density after varying times $\Delta t\omega_\perp=0.9, 1.23, 1.32, 1.41, 1.50,$ and $1.71$ of in-trap evolution after switching off particle interactions and (g-o) show the corresponding spatial phases. The salmiakki shape highlighted in frames (p-r) which are obtained by holding the condensate in-trap for $\Delta t\omega_\perp=1.02$ followed by $\tau \omega_\perp=0.29, 0.31, 0.35$ ballistic evolution, emerges between the two line foci visible in (b) and (e). The phase singularities at the cores of the quantized vortices and antivortices, which form the skeleton supporting the caustic, are marked by red and white filled circles. The field of view of the frames (a-n) is $12\times12$ in units of $a_{\rm osc}$ and $6\times4$ for (p-r). Movies corresponding to these frames are shown in the Supplemental Material \cite{Supplement}.} 
\label{figS1}
\end{figure*}

\subsection{Pentafoil aberration}
In the case of five-fold symmetric lens aberration, pentafoil $h$-wave snow flake patterns shown in Fig.~\ref{fig5} emerge. In this case $V_{\rm lens}({\bf r})= 0.0001( 5x^4y -10x^2y^3 +y^5) \hbar\omega_\perp/a_{\rm osc}^5\propto {\mathcal Z}_5^5$ and $\Delta t\omega_\perp=0.0075$. Frames (a)-(d) are obtained using method (II) and the frames (e) and (f) are for the case (I). It is straightforward to extend these results to include arbitrary lens aberrations which yield a hierarchy of ever more complex diffraction detail and vortex skeletons. Irregular focusing can be modeled by considering a stochastic lens aberration obtained by randomly choosing the coefficients $\beta(n,m)$ of the lens polynomials. Alternatively, a speckled lens aberration can be produced for example by Fourier transforming a band limited white noise random potential. 

\subsection{Astigmatism aberration induced by anisotropic harmonic oscillator potential}

Previously, we considered aberrated atom-optic matter wave lensing for toroidal condensates. Such trapping geometry is ideally suited for such studies because the high degree of cylindrical symmetry achievable in Laguerre--Gauss laser potentials minimizes aberrations due to external potentials. In addition, by removing the central condensate density, the fine structure and visibility of the underlying caustic is enhanced. However, the generic caustic patterns are also generated in the case of more conventional harmonically trapped quantum gases. Figure \ref{figS1} (see also the supplemental movies 7 and 8) shows snapshots of an astigmatism induced caustic structure developing in a Bose-Einstein condensate trapped in an anisotropic harmonic potential with Cartesian frequencies $\omega_x/\omega_\perp =1$, $\omega_y/\omega_\perp =1.3$, $\omega_z/\omega_\perp =1.6$. The interaction strength is suddenly reduced from $g'_i=5000$ to $g'_f=0$ and the system is allowed to evolve in-trap. The top row (a-f) shows the resulting condensate density for in-trap evolution times, $\Delta t\omega_\perp=0.9, 1.23, 1.32, 1.41, 1.50,$ and $1.71$, respectively and the middle row (g-o) shows the corresponding spatial phase patterns, respectively.

\begin{figure*}
\includegraphics[width=2\columnwidth]{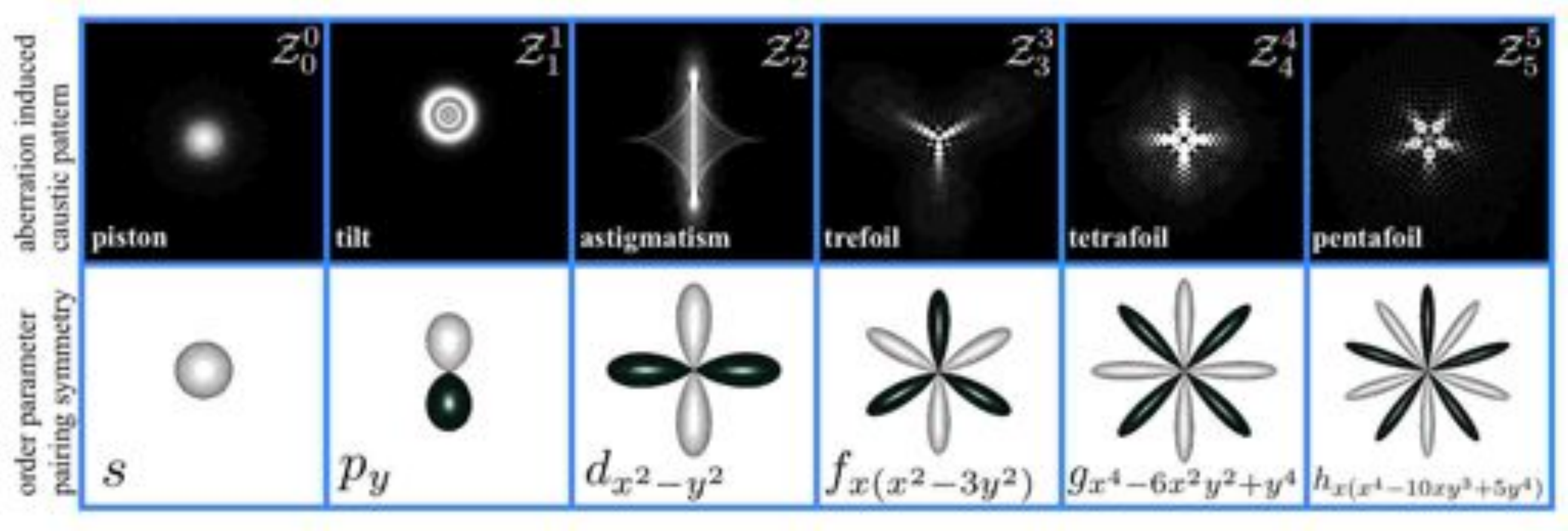}
\caption{(Color) Caustic patterns (top row) produced by matter wave aberrations defined by the respective Zernike polynomials $\mathcal{Z}^m_n$ corresponding to the order parameter symmetries illustrated in the row below. The images of the caustics are calculated using the diffraction integral \cite{Petersen2013a} and the symmetries of the partial wave interactions are visualized using spherical harmonics.} 
\label{fig7}
\end{figure*}

The significant $d_{x^2-y^2}$ symmetric quadrupolar anisotropy $\propto{\mathcal Z}_2^2$ of the trap causes astigmated lensing of the atoms. The sequence of images Fig.~\ref{figS1} (p-r) highlighting the salmiakki shaped caustic is obtained by first evolving the condensate in-trap for $\Delta t\omega_\perp=1.02$ followed by $\tau \omega_\perp=0.29, 0.31, 0.35$ ballistic evolution, respectively. This telltale signature of lens astigmatism can be compared with similar patterns in Ref.~\cite{Petersen2013a} where images of the same caustic were observed in an electron microscope, where the larger momentum of the electrons yields finer diffraction detail inside the caustic. On astigmatic focussing the matter wave passes two narrow line foci (b) and (e) between which the characteristic astigmatic caustic structure is observable. The ghost vortices and antivortices in the four quadrants of the frames in Fig.~\ref{figS1} are an inherent feature of the astigmatic caustic. 

\subsection{Diffraction catastrophes and Bose-novae}
As demonstrated in Sec. III. E. an anisotropic harmonic trap can act as an astigmatic lens when the strength of the repulsive interparticle interactions is changed, causing the atoms to be channelled toward the focal volume of the trap. Similarly, the dipole-dipole interactions possess the symmetry of the astigmatic lens perturbation and produce the same quadrupolar anisotropy (intrinsic ${\mathcal Z}_2^2$ astigmatism) to the condensate density as an external harmonic trap anisotropy. Therefore both external anisotropic harmonic trapping or internal dipolar interactions can be used as a source for the universal astigmatic ``d-wave/cloverleaf/rhomboid/salmiakki" caustic patterns. 

The correspondence between Bose-novae phenomena and caustic formation can be quantified by considering the density profile of a ground state Thomas--Fermi condensate $n({\bf r}) = (\mu - V_{\rm trap}({\bf r}))/g_i$ in the transverse $x_1-x_2$ plane perpendicular to the imaging axis in an anisotropic harmonic trap $V_{\rm trap}= \frac{1}{2}m\omega^2(\lambda^2_1 x_1^2+\lambda_2^2 x_2^2)$, where $\mu$ is the chemical potential, $\omega=(\omega_1+\omega_2)/2$, $\lambda_p=\omega_p/\omega$, $g_i$ is the initial value of the $s$-wave coupling constant. Upon changing the strength of the mean-field from $g_i$ to $g_f$, the resulting lens potential is 
\begin{eqnarray}
V^{\rm trap}_{\rm lens} &=& V_{\rm trap} + g_f n  \notag\\
&=&\alpha_1{\mathcal Z}_0^0+\alpha_2{\mathcal Z}_2^0+\alpha_3{\mathcal Z}_2^2,
 \end{eqnarray} 
where the constants are 
$\alpha_1 = g_f\mu/g_i$,
$\alpha_2 = \frac{1-g_f/g_i}{4}(\lambda_1^2+\lambda_2^2)m\omega^2$,
and
$\alpha_3 = \frac{1-g_f/g_i}{4}(\lambda_1^2-\lambda_2^2)m\omega^2$, corresponding to the respective coefficients of the piston, defocus and astigmatism. 

Within the Thomas--Fermi radius the dipole-dipole interaction adds an interaction induced lens potential $\Phi_{dd}(r)=e_{dd }m\omega^2(1-3\cos^2(\theta))r^2/5$, where $e_{dd}$ is the ratio of dipolar length and s-wave scattering length \cite{Lahaye2009a}. The corresponding transverse lens potential is therefore
\begin{eqnarray}
V^{\rm dipole}_{\rm lens} &=&\beta_1{\mathcal Z}_0^0+\beta_2{\mathcal Z}_2^0+\beta_3{\mathcal Z}_2^2,
\end{eqnarray} 
where the constants are 
$\beta_1 = 0$,
$\beta_2 = -e_{dd }m\omega^2/10$,
and
$\beta_3 = 3e_{dd }m\omega^2/10$. 
Hence, the dipole-dipole interactions can yield the same ${\mathcal Z}_2^2$ astigmatism lens aberration as anisotropic harmonic trapping resulting in the same universal astigmatic caustic. The ratio $R=|\alpha_3/\beta_3| = |5(1-g_f/g_i)(\lambda_1^2-\lambda_2^2)/6e_{dd }|$ can be used to estimate the relative strengths of the trap astigmatism and dipolar astigmatism. It is interesting to calculate this quantity for the conditions in the experiments \cite{Donley2001a,Lahaye2008a,Aikawa2012a} for which we obtain, by setting $e_{dd }=1$, the respective values $7.8, 0.39$, and $0.23$. This suggests that the main contribution to the astigmatism would be due to the trap anisotropy $R>1$ in the case of Ref.~\cite{Donley2001a} and due to dipolar interactions $R<1$ in Refs~\cite{Lahaye2008a} and \cite{Aikawa2012a}. Comparison of Figs~\ref{figS1}(p-r) with Figure 5(f) in Ref.~\cite{Donley2001a}, Figure 1(d) in Ref.~\cite{Lahaye2008a}, and Figure 3(e) in Ref.~\cite{Aikawa2012a} further supports our interpretation of such Bose-novae in terms of aberrated matter wave lensing.

\subsection{Applications of matter wave diffraction catastrophes}

Here we speculate on the potential uses of the presented singular nonlinear atom-optics protocols. Recently, it has become experimentally possible to synthesize higher order partial wave interactions in cold atom systems \cite{Williams2012a}. The $d$-wave symmetry of the long-range dipole-dipole interactions has already been detected by imaging the telltale caustic shapes produced by lens astigmatism in Bose--nova experiments \cite{Lahaye2008a,Aikawa2012a}. More generally, measuring the universal caustic patterns produced by focusing matter waves might enable the detection of the pairing symmetry of the order parameters of cold bosonic, fermionic and molecular quantum gases. For example, microscopic $f$-wave pairing interactions could be revealed on macroscopic scales by a trefoil pattern of the elliptic umbilic diffraction catastrophe, whereas the pentafoil caustic reflects the underlying $h$-wave symmetry. Notice also that the vortex skeletons of the caustics act as fiducial markers mapping out the symmetry of the aberrating potential. The correspondence between lens aberrations described in terms of Zernike polynomials and partial wave interactions described in terms of spherical harmonics is illustrated in Fig.~\ref{fig7}. The generic microscopic particle interactions encapsulated in the potential $U({\bf r- r'},t)$ result in a mean-field lens potential
\begin{eqnarray}
V_{\rm lens}^{\rm int}({\bf r},t) &=& \int U({\bf r- r'},t)|\psi({\bf r'},t)|^2d^3{\bf r'} \notag \\
                                      &=& \sum_{n=0}^\infty \sum_{l=0}^n \sum_{m=-l}^l\beta(n,m,l,t) {\mathcal Z}_{nl}^m(\bf r),
\label{linssi}
\end{eqnarray}
which can be expressed in terms of 3D Zernike functions ${\mathcal Z}_{nl}^m(\bf r)$, with $n-l$ even \cite{Canterakis1999a,Novotni2003a,Kihara2011a}.

Replacing the diffraction catastrophes of matter waves by those made of light opens the possibility to create exotic and steep walled dark field light shift potentials for cold atoms since the tuneable diffraction catastrophes of the laser field can be made static in the reference frame of the trapped ground state condensate. Bragg scattering with Laguerre--Gauss beams has been used for transferring orbital angular momentum, or the phase of the photon field, to Bose--Einstein condensates \cite{Andersen2006a,Simula2008a}. Other kinds of spatial gradients of light-induced momentum have also been used to create artificial gauge fields for ultracold atoms \cite{Lin2009a,Lin2011a}. Similarly, photon fields sourced from aberration induced diffraction catastrophes may enable the creation of novel topological quantum states of ultracold atoms \cite{Cooper2011a,Simula2012a,Cooper2013a}. Creating such optical flux beams carrying quantized vortices with intervortex spacing on the order of the wavelength of the light only requires a single beam of light to be passed through a suitable aberrating (programmable) lens instead of interfering beams from multiple laser sources. An ability to systematically imprint staggered vortex lattice states in Bose--Einstein condensates would be particularly useful for studies of quantum turbulence. Atom-optic diffraction catastrophes are robust possessing intrinsic structural and topological stability. It has been predicted that time-of-flight experiments could be used for detecting topologically ordered quantum states in cold atom systems \cite{Alba2011a}. It will be interesting to investigate whether states with topological order also leave a distinct observable signature to matter wave caustics.

The matter wave diffraction catastrophes are generic. In spinor condensates the caustic vortex skeletons will be composed of fractional vortices and may result in non-Abelian vortex networks \cite{Kobayashi2009a,Huhtamaki2009a,Ruben2010a}. It should also be possible to use the method of atom-optic lens aberrations and focussing to nucleate vortex rings \cite{Petersen2013a} and knotted matter wave vortices in quantum gas systems.
\\

\section{Conclusions}

We have presented a singular atom-optics protocol to create and observe aberration induced diffraction catastrophes and the associated skeletons of quantized vortices using cold atom experiments. Imaging resolution of the internal structure of such caustics could be boosted by adding a defocus (antitrapping) to the atom-optical lens perturbation to increase the physical size of the diffraction catastrophes sufficiently to allow resolving the fine structure of the diffraction catastrophes. The atom-optical lens aberrations can be produced either by external potentials used for trapping and manipulating the atoms or intrinsically by internal particle interactions. In particular, both anisotropic harmonic trapping and dipolar interactions can result in the same astigmatic aberration due to their quadrupolar symmetry. We have discussed potential applications of diffractive singular nonlinear matter wave optics experiments suggesting that they could be useful for detecting the microscopic pairing interactions of the order parameter. 


\end{document}